\documentclass[aps,prb,showpacs,twocolumn,groupedaddress,superscriptaddress,amsmath,amssymb]{revtex4}
\usepackage{amsmath} \usepackage{amssymb} \usepackage{graphicx}
\usepackage{epstopdf} \usepackage{bm}

\begin{document}

\title{Spin and energy relaxation in germanium studied by
  spin-polarized direct-gap photoluminescence}

\author{Fabio Pezzoli} \thanks{The first two authors have equal
  contribution to this work.}  \email{fabio.pezzoli@unimib.it}
\affiliation{LNESS-Dipartimento di Scienza dei Materiali,
  Universit\`{a} degli Studi di Milano-Bicocca, via R. Cozzi 53,
  I-20125 Milano, Italy}

\author{Lan Qing} \thanks{The first two authors have equal
  contribution to this work.} \email{fabio.pezzoli@unimib.it}
\affiliation{Department of Physics and Astronomy, University of
  Rochester, Rochester, New York 14627}

\author{Anna Giorgioni} \affiliation{LNESS-Dipartimento di Scienza dei
  Materiali, Universit\`{a} degli Studi di Milano-Bicocca, via
  R. Cozzi 53, I-20125 Milano, Italy}

\author{Giovanni Isella} \affiliation{LNESS-Dipartimento di Fisica del
  Politecnico di Milano, Polo di Como, via Anzani 42, I-22100 Como,
  Italy}
	
\author{Emanuele Grilli} \affiliation{LNESS-Dipartimento di Scienza
  dei Materiali, Universit\`{a} degli Studi di Milano-Bicocca, via
  R. Cozzi 53, I-20125 Milano, Italy}

\author{Mario Guzzi} \affiliation{LNESS-Dipartimento di Scienza dei
  Materiali, Universit\`{a} degli Studi di Milano-Bicocca, via
  R. Cozzi 53, I-20125 Milano, Italy}

\author{Hanan Dery} \affiliation{Department of Physics and Astronomy,
  University of Rochester, Rochester, New York 14627}
\affiliation{Department of Electrical and Computer Engineering,
  University of Rochester, Rochester, New York 14627}

\begin{abstract}
  Spin orientation of photoexcited carriers and their energy
  relaxation is investigated in bulk Ge by studying spin-polarized
  recombination across the direct band gap. The control over
  parameters such as doping and lattice temperature is shown to yield
  high polarization degree, namely larger than 40\%, as well as a
  fine-tuning of the angular momentum of the emitted light with a
  complete reversal between right- and left-handed circular
  polarization. By combining the measurement of the optical
  polarization state of band-edge luminescence and Monte Carlo
  simulations of carrier dynamics, we show that these very rich and
  complex phenomena are the result of the electron thermalization and
  cooling in the multi-valley conduction band of Ge.  The circular
  polarization of the direct-gap radiative recombination is indeed
  affected by energy relaxation of hot electrons via the $X$ valleys
  and the Coulomb interaction with extrinsic carriers. Finally,
  thermal activation of unpolarized $L$ valley electrons accounts for
  the luminescence depolarization in the high temperature regime.
\end{abstract}


\maketitle


\section{Introduction}

The coupling between the angular momentum of photons and the spin
angular momentum of carriers, termed optical
orientation,\cite{Dyakonov_OO, Zutic04} has been so far recognized as
one of the main challenges in group IV materials.\cite{Zutic04,
  Zutic06, Li10} The poor absorption and emission efficiencies,
associated to the weak electric-dipole transitions of the fundamental
indirect-gap of such semiconductors, jeopardized the optical
exploitation of their rich spin physics. Indeed, silicon and germanium
feature lattice inversion symmetry and predominant spin-less
isotopes,\cite{Zutic04, Fodor06} yielding long spin coherence
times.\cite{Fodor06,Appelbaum07} This feature is crucial for the
effective implementation of spintronic devices,\cite{Wolf01} and
quantum information processing.\cite{Zutic04, Fodor06, Wu10}

Only very recently, however, the quasi-direct behavior of Ge has
sparked interest in its photonic properties,\cite{Liu10, Liang10} and
stimulated the use of various optical schemes aiming at addressing its
spin physics.\cite{Loren09, Virgilio09, Guite11, Pezzoli12} In Ge, the
absolute minimum of the conduction band (CB) is at the $L$~point of
the Brillouin zone, but there exists a local minimum at the
zone-center $\Gamma$.  At low temperature the former leads to an
indirect energy gap of 0.744~eV,\cite{Zwerdling59} the latter to a
direct-gap of 0.898~eV.\cite{Zwerdling59} Optical orientation via
absorption of circularly polarized photons at the direct energy gap
can then be exploited to readily investigate the spin properties of
Ge.\cite{Rioux10,Loren11,Li12b,Li13} Noticeably, spin-polarized
electrons optically pumped in the $\Gamma$ valley can conserve their
spin during ultrafast scattering to the lower energy
$L$ valleys,\cite{Pezzoli12} which are responsible for charge and spin
transport. Such mechanism, which is absent in the widely studied III-V
semiconductors, makes Ge spin dynamics very rich, but still poorly
understood. In addition, it provides a viable approach to inject
spin-polarized carriers in Ge without relying on ferromagnetic gate
stacks, which are prone to low efficiency and experimental artifacts
due to defects and localized interfacial states.\cite{Uemura12,
  Jain12}

Optical orientation has indeed recently allowed progress in the
investigation of spin dynamics of both electrons and holes in bulk
Ge. Spin relaxation times of holes have been shown to be below
1~ps,\cite{Loren11} whereas electron spin relaxation times are in the
ns range below 200~K.\cite{Guite11, Guite12, Hautmann12} Yet very
little is known about spin flip scattering by dopants and the role
played by impurities in determining spin dynamics in different
temperature regimes.\cite{Song12,Restrepo12}

Inspired by the quasi direct-gap behavior of Ge and by the possibility
to optically initialize spins, we report in this work a spin-polarized
photoluminescence (PL) study focused on the recombination across the
direct-gap of bulk Ge over a wide doping and temperature range. We
directly measured the polarization state of the direct-gap emission by
means of Stokes analysis,\cite{Goldstein_00} shedding light on the
optical orientation process and on the interplay between energy and
spin relaxation channels. The role of carrier thermalization and
cooling in determining luminescence polarization in Ge are disclosed
by Monte Carlo simulations, in which non-equilibrium kinetics of
photoexcited electrons is used to extract the circular polarization
degree of the emitted light under steady state conditions.

Our analysis shows that in Ge the state of light polarization of the
direct-gap emission is governed by kinetics of spin-polarized
electrons, highlighting the role of energy relaxation of hot electrons
within the $X$ valleys. The direct comparison between theory and
experiments points out that above 170~K back-scattering of unpolarized
electrons from $L$ to $\Gamma$ valley leads to a decay of the
circular polarization degree of direct-gap emission, no matter the
doping of the bulk samples. Remarkably, at temperatures below 170~K a
complete reversal of the helicity of light polarization can be
obtained either by changing the doping level or the lattice
temperature. In addition, a maximum in the polarization degree of the
emitted light is obtained around 100-150~K, reaching in intrinsic Ge
samples surprisingly high values, i.e. larger than $40\%$.

Such puzzling behavior stems for the complex carrier dynamics taking
place in the multi-valley band structure of Ge, where the electron
spin (and ensuing direct-gap luminescence polarization) is dictated by
cooling of hot electrons via Coulomb collisions and intervalley
scattering between $\Gamma$, $X$ and $L$ valleys.

The paper is organized as follows. In Sec.~\ref{sec:exp} we discuss
the samples and their optical properties, extending the results to the
polarimetric analysis of the PL. In Sec.~\ref{sec:theo} we describe in
detail the Monte Carlo simulations. These calculations unravels energy
and spin relaxation channels, finally disclosing their relative
contribution in determining the experimental findings. After having
discussed in Sec.~\ref{sec:physpic} the physics underlying the
observed phenomena, we then in Sec.~\ref{sec:con} summarize the
results and provide the future perspectives of this work.

\section{Experiments}
\label{sec:exp}
\subsection{Experimental Details}

We studied a set of bulk Ge(001) samples having a different type and
level of doping: (i) a $p$-type Ge:Ga wafer, with an acceptor
concentration of $3.6 \times 10^{18}$ $\mathrm{cm^{-3}}$, named
\textit{p}$^+$-Ge, (ii) a $n$-type Ge:As, $6^\circ$ miscut, with a
donor concentration of $8.3 \times 10^{16}$ $\mathrm{cm}^{-3}$, named
\textit{n}-Ge, (iii) a $p$-type wafer with a doping concentration of
$1.4 \times 10^{15}$ $\mathrm{cm}^{-3}$, named \textit{p}$^-$-Ge, and
(iv) an intrinsic Ge sample, with a resistivity of $47$~$\Omega$ cm,
named \textit{i}-Ge. The doping levels have been obtained by means of
room temperature resistivity measurements. Sample characteristics are
summarized in Tab.~\ref{tab:samples}.

PL measurements were carried out in back-scattering geometry in the
temperature range between 4 and 300~K. The samples were excited by a
$\mathrm{Nd:YVO_4}$ laser. The excitation energy was 1.165~eV and the
light was left-handed circularly polarized ($\sigma^{-}$). The laser
spot size on the sample surface was about 100 $\mathrm{\mu m}$,
resulting in a power density of $\sim 1$~$\mathrm{kW/cm^2}$. The
polarization state of the luminescence was probed by a Stoke analyzer,
i.e. an optical retarder followed by a linear polarizer. Hereafter we
define the analyzer angle as the angle determined by the optical axis
of the polarizer and that of the retarder. In the experiments the
analyzer angle spans 360$^\circ$ with a resolution of 1$^\circ$. PL
was dispersed by a spectrometer equipped with a
thermoelectrically-cooled InGaAs array detector, with a cut-off
starting at about 0.755~eV. The energy accuracy was $\sim 4$ meV. The
multiple-channel detector measured the amplitude of the PL spectra as
a function of the analyzer angle. The analysis of the peak amplitude
modulation provided the Stokes parameters, $\mathrm{S_i}$, with
$i$=0-3, which permit the full characterization of the polarization
state of light.\cite{Goldstein_00,Collett_00,Pezzoli12} Since
partially polarized light can be considered as a superposition of
unpolarized and completely polarized light, we can define the degree
of polarization, $\rho$, as:\cite{Goldstein_00, Collett_00}
\begin{equation}
  \label{eq:rho}
  \mathrm{\rho = \pm \frac{\sqrt{S_1^2+S_2^2+S_3^2}}{S_0}}
\end{equation}

For circularly polarized light, the sign of the polarization degree
has been chosen to be consistent with the sign of the Stokes parameter
$\mathrm{S_3}$, which defines whether the light is left-handed
($\sigma^-$): $\mathrm{-1\leq S_3 < 0}$, or right-handed ($\sigma^+$):
$\mathrm{0 < S_3\leq 1}$.

\begin{table}\caption{Ge bulk samples investigated in this work.}
  \vspace{0.1cm}
  \label{tab:samples}
  \tabcolsep=0.49cm
  \renewcommand{\arraystretch}{1.5}
  \begin{tabular}{c|cc}
    \hline \hline
    Sample &  resistivity ($\Omega$ cm) & doping ($\mathrm{cm^{-3}}$) \\ \hline \hline
    $p^+$ & 0.0046 & $3.6\times 10^{18}$ \\
    $\!n$ & 0.358 & $8.3\times 10^{16}$ \\
    $p^-$ & 2.39 & $1.4\times 10^{15}$ \\
    $\!\!i$ & 47 & $\approx 10^{13}$ \\ \hline \hline
  \end{tabular}
\end{table}

\subsection{Photoluminescence}
\label{subsec:PL}

\begin{figure*}
  \centering
  \includegraphics[width=17.8cm]{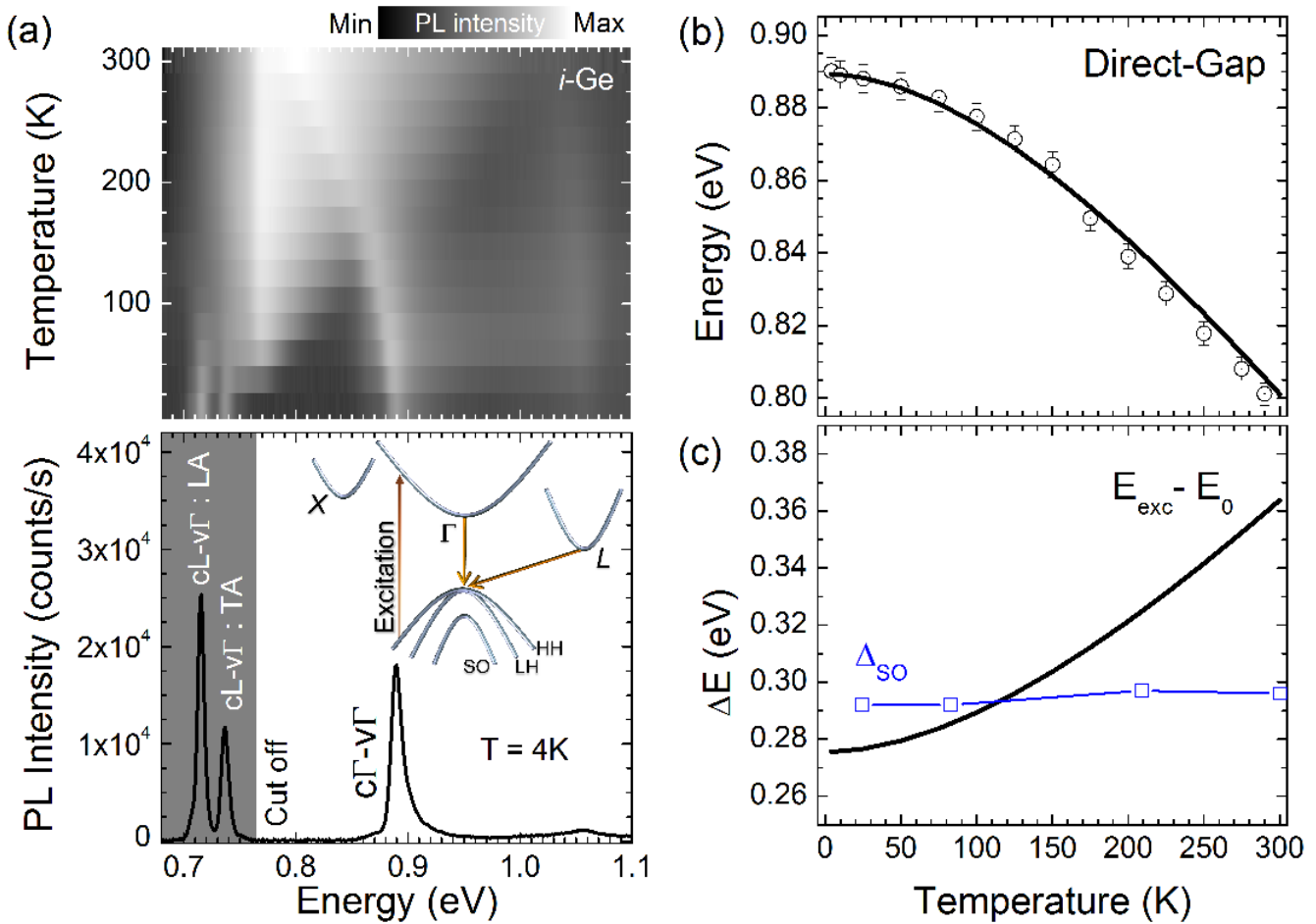}
  \caption{(a) Upper panel: Color-coded map of the PL intensity
    describing the temperature dependence of the photoluminescence
    (PL) spectra of \textit{i}-Ge. Lower panel: 4~K PL data. Direct
    gap, $\mathrm{c}\Gamma-\mathrm{v}\Gamma$, and indirect-gap
    transitions, $\mathrm{c}L-\mathrm{v}\Gamma$, mediated by
    longitudinal (LA) and transverse (TA) acoustic phonons are
    indicated along with the detector cut off. The inset shows the
    physical processes underlying the luminescence mechanisms. (b)
    Energy variation of the direct-gap with the temperature. Open dots
    are the PL peak positions measured in \textit{i}-Ge, whereas the
    solid line represents the Varshni's law according to the
    parameters of Ref.~[\onlinecite{Varshni67}]. (c) Temperature
    dependence of the total excess energy, $\mathrm{\Delta E}$, for
    carriers photoexcited in Ge by a laser energy at 1.165~eV (solid
    black line).  The blue squares show the variation of the energy of
    the split-off band, $\mathrm{\Delta_{SO}}$, with temperature
    according to Ref.~[\onlinecite{Nishino69}].}
  \label{fig:1}
\end{figure*}

The PL spectra of bulk, undoped Ge(001) are outlined in
Fig.~\ref{fig:1}(a), along with a schematics of the radiative
recombination processes.

At 4~K, the spectral feature observed at 0.885~eV is the direct band
gap emission, which is due to the recombination of holes with
electrons, directly photo-generated within the $\Gamma$ valley by the
laser excitation at 1.165~eV. The lifetime of $\Gamma$ electrons is
dominated by their fast scattering out of the optically coupled region
towards the side $X$ and $L$ valleys and limited to few hundreds of
fs.\cite{Zhou94, Mak94, Bailey95, Kolata12} The relaxation processes
towards the edge of the $\Gamma$ valley are significantly slower than
in typical III-V materials, where phonon scattering is driven by the
efficient Fr\"ohlich interaction, which is absent in Ge since the
crystal bonds are purely covalent. In addition, the deformation
potential interaction between long-wavelength optical phonons and CB
electrons is weak in Ge due to the space inversion
symmetry.\cite{YuCardona_OO} As a result, scattering to the
side-valleys is more efficient than the intravalley cooling at
$\Gamma$.

The majority of electrons excited in the $\Gamma$~valley will thus
reach and dwell in the $L$~valley, experiencing a relatively long
lifetime $\tau_L$ between tens and hundreds of $\mathrm{\mu}$s at room
temperature.\cite{Pankove71} The two main features shown in the lower
panel of Fig.~\ref{fig:1}(a) at 0.737~eV and 0.715~eV can thus be
ascribed to recombination between holes at $\Gamma$ and $L$-electrons
mediated by transverse (TA) and longitudinal (LA) acoustic phonons,
respectively.\cite{Lieten12} By increasing the temperature these two
PL bands redshift, leaking out of the spectral response range of the
detector except for their high energy tail [see upper panel of
Fig.~\ref{fig:1}(a)].
The indirect-gap emission will not be discussed further, since in this
work we focus on the spin and energy relaxation of electrons directly
photoexcited in the $\Gamma$ valley.
Finally, we point out that the weak feature observed in
Fig.~\ref{fig:1}(a) at about 1.056 eV is almost temperature
independent and can be ascribed to resonant inter-valence-band Raman
scattering \cite{Wagner84, Tanaka94}.

As shown in Fig. \ref{fig:1}(a) by the color-coded map of the PL
intensity, the direct-gap emission redshifts with increasing the
lattice temperature, as a result of the band gap shrinkage. Above
125~K it is superimposed to the high energy tail of the indirect-gap
PL. The observed peak position of the direct band gap emission is in
good agreement with the temperature dependence of the gap as described
by the Varshni's law,\cite{Varshni67} reported as a solid line in
Fig.~\ref{fig:1}(b).

By sweeping the sample temperature, while keeping fixed the excitation
energy $\mathrm{E_{exc}}$, we gather spectroscopic access to high
energy states within the bands. Holes (electrons) are excited in the
VB (CB) with a total excess energy $\mathrm{\Delta E =
  E_{exc}-E_{0}}$, which increases with the temperature (see
Fig.~\ref{fig:1}(c)). $\mathrm{E_{0}}$ is the energy of the direct
gap. It is worth noticing that already in the low temperature range,
the excess energy for CB electrons is comparable to the energy
difference of the $\Gamma$ and $X$ valley bottoms, $\delta
\epsilon_{x,\Gamma} \approx= 0.04$~eV, thus activating scattering to
the side $X$ valleys as an energy relaxation channel. Moreover, the
threshold for optical transitions involving the split-off band (SO),
lying at $\mathrm{\Delta_{SO} \sim 0.29}$~eV below the VB
edge,\cite{Nishino69} is approached at about 125~K. At low
temperatures electrons can indeed be photo-generated from the SO band
directly at the bottom of the CB, whereas the vast majority of high
energy electrons, promoted far above the CB edge, result from optical
transitions from heavy hole (HH) states at the top of the VB. The
oscillator strength for transitions at $k=0$ involving HH states is
three times the one involving light hole (LH) states.

Hereafter, we focus on the samples with the lowest and the highest
doping level: \textit{i}-Ge and \textit{p}$^+$-Ge, respectively, to
pin down the impact of doping on the PL spectra. At a fixed
temperature and as impurities are introduced in the Ge host crystal,
the direct-gap shifts to lower energies, the shift being larger for
samples with a larger doping level.\cite{Haas62, Wagner84}

The PL analysis can reveal important features related to the different
phenomena taking place on the photoexcited carriers in doped
samples. Indeed the spectral dependence of the direct-gap PL is the
result of the joint density of states weighted by the distribution
function of carriers, finally leading to a skewed lineshape. We thus
modeled the lineshape of the PL spectra by using the following
exponentially modified Gaussian distribution, i.e. the convolution of
an exponentially decaying function and a normal distribution
\cite{Grande07}:
\begin{equation}
  \label{eq:modgauss}
  \begin{split}
    \mathrm{\Phi(E)}=\;&\mathrm{\frac{A}{\lambda}\exp\left[\frac{1}{2}\left(\frac{w}{\lambda}\right)^2-\frac{E-E_{max}}{\lambda}\right]}\\
    &\times\mathrm{\int^z_{-\infty}\frac{1}{\sqrt{2\pi}}\exp \left(-\frac{x^2}{2}\right)dx}\,,
  \end{split}
\end{equation}
where
\begin{equation}
  \label{eq:z}
  \mathrm{z=\frac{E-E_{max}}{w}-\frac{w}{\lambda}}
\end{equation}
$E$ is the photon energy, $\mathrm{E_{max}}$ is the PL peak position,
$A$ is the amplitude of the PL band. $w$ is the width of the Gaussian
component, which allows us to estimate inhomogeneous broadening
effects, while $\lambda$ is the modification or skewness factor, which
quantifies the asymmetry in the lineshape due to the thermal
distribution of carriers in the band. The latter two parameters thus
provide valuable information about the population of the carriers
which experience radiative recombination from the $\Gamma$ valley.

\begin{figure}
  \centering
  \includegraphics[width=8.6cm]{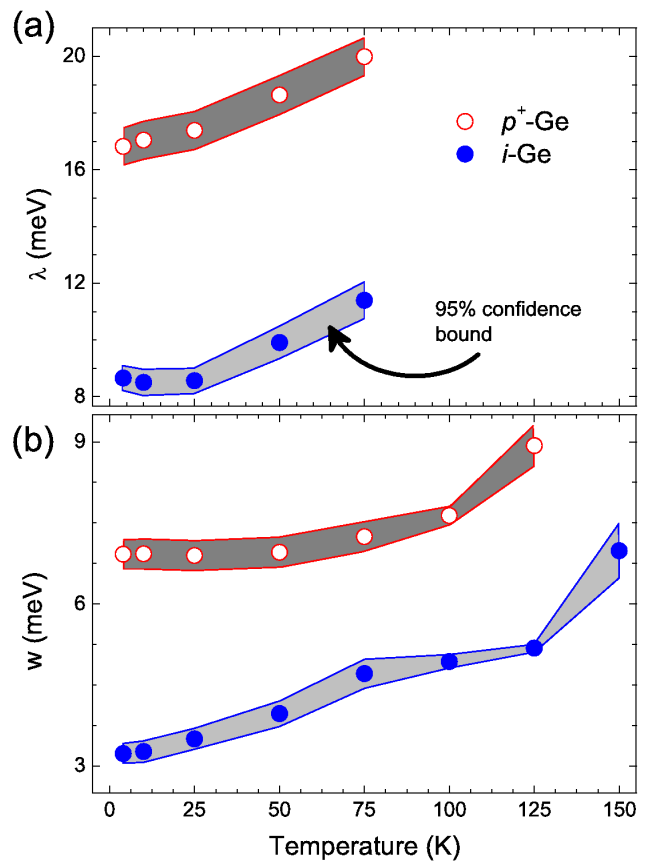}
  \caption{Temperature dependence of the modification factor,
    $\lambda$, and linewidth, $w$, of the direct-gap PL of $p^+$-Ge
    (open dots) and \textit{i}-Ge (full dots) samples. The shadowed
    gray area defines the 95\% confidence bound region as obtained
    from the fitting of the spectra with the exponentially modified
    Gaussian function reported in Eq.~(\ref{eq:modgauss}).}
  \label{fig:2}
\end{figure}

The results of the modeling of the experimental data according to
Eq.~(\ref{eq:modgauss}) are summarized in Fig.~\ref{fig:2} highlighting
$\lambda$ and w in the low temperature range, i.e. where the lineshape
analysis is less affected by the tail of the indirect-gap
emission. Noticeably, the average carrier temperature, elucidated by
the skewness factor shown in Fig.~\ref{fig:2}(a), increases sharply
above $\approx 50$~K in the investigated samples. We therefore expect
hot carrier phenomena and carrier-carrier interactions to play a major
role in determining spin relaxation for temperatures above $\approx
50$~K. Finally, as expected from band filling effects,
Fig.~\ref{fig:2}(b) demonstrate that the PL linewidth increases with
the concentration of extrinsic carriers.

\subsection{Polarization-Resolved Photoluminescence}

\begin{figure}
  \centering
  \includegraphics[width=8.6cm]{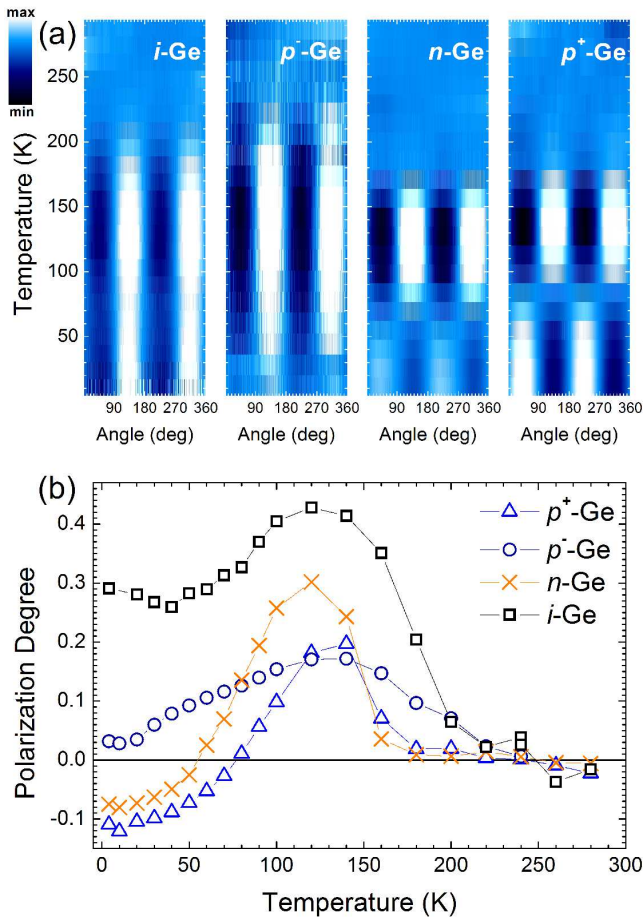}
  \caption{(a) Modulation of the peak intensity with the analyzer
    angle measured in \textit{i}-Ge, \textit{p}$^-$-Ge, \textit{n}-Ge
    and \textit{p$^+$}-Ge as a function of temperature. The minima
    (dark) and maxima (bright) are visible from the color-coded
    scale. (b) Experimental data for the polarization degree of the
    studied samples as a function of doping and temperature. }
  \label{fig:3}
\end{figure}

To address optical orientation of spins and their depolarization
channels in Ge, we measured the polarization state of the direct-gap
emission by means of the Stokes
analysis.\cite{Goldstein_00,Collett_00,Pezzoli12}

Zone center valence band (VB) states transform with the same symmetry
operations of atomic $p$ orbitals and are characterized by a total
angular momentum quantum number $J$=3/2 (for HH and LH) and $J$=1/2
(for SO states). The projection of the total angular momentum along
the quantization axis, conveniently chosen to be parallel to the
angular momentum of the exciting light beam, is $\mathrm{J_z=\pm3/2}$
for HH, $\mathrm{J_z=\pm1/2}$ for LH and SO band. CB states at
$\Gamma$ are $s$-like and their total angular momentum coincide with
their spin and are labeled by $J$=1/2 and $\mathrm{J_z=\pm1/2}$,
i.e. $|J,\mathrm{J_z}\rangle = |1/2, \pm1/2\rangle$.

In a bulk material under external illumination with left-handed
($\sigma^-$) circularly polarized light, both final CB states
$\mathrm{J_z=+1/2}$ and $\mathrm{J_z=-1/2}$ are populated, with a
relative weight which depends upon the excitation
energy,\cite{Rioux10} and the strength of the dipole allowed
transitions involved in the absorption process. The photo-generated
spin-polarized carriers will then diffuse within the sample, possibly
experiencing spin relaxation prior to radiative recombination. Spin
relaxation mechanisms tend to equalize spin up and spin down
populations. In bulk material, spin relaxation for holes is faster
than for electrons because of the strong mixing of the HH and LH
states.\cite{Zutic04,Hautmann11} As a result, under steady state
conditions, the non-equilibrium electron spin population will govern
the helicity and polarization degree of the direct-gap
emission. Studies of the circular polarization of luminescence have
allowed the investigation of spin physics of non-equilibrium carriers
in direct-gap semiconductors such as GaAs, whereas no detailed studies
have been reported so far for bulk Ge.\cite{Zutic04} The similarity
between the band structure of Ge and of III-V compounds near the
center of the Brillouin zone lead to analogous optical orientation
during the absorption process. \cite{Rioux10} On the other hand, the
ultrafast lifetime of $\Gamma$ valley electrons in Ge, and the role
played by scattering to the side valleys can possibly lead to crucial
differences in the spin relaxation channels and eventually in the
luminescence polarization of direct-gap radiative recombination.

Fig.~\ref{fig:3}(a) shows the contour plot of the direct-gap PL
intensity as a function of the angle of the polarization analyzer at
temperatures ranging from 4~K to room temperature.  We can identify
three regimes.

In the low temperature range, i.e. below $\sim 90$~K, for all the
samples the amplitude modulation of the PL peaks reveals a sinusoidal
behavior. According to Stokes analysis, this is the fingerprint of
circularly polarized light and thus of optically oriented electron
spins. Indeed the lifetime, $\tau$, for electrons lying in the
$\Gamma$ valley of Ge is limited by the ultrafast scattering to the
side valleys being hundreds of fs.\cite{Zhou94, Mak94} This process is
much shorter than the spin relaxation time of CB
electrons, $\tau_{es} \sim$ ns,\cite{Guite11,Hautmann12,Guite12,Pezzoli12} finally yielding
circularly polarized luminescence. We recall that the observed
luminescence polarization degree is $\rho \propto
(1+\tau/\tau_{es})^{-1}$.\cite{Parsons69}

Remarkably, in both \textit{p}$^+$-Ge and \textit{n}-Ge the direct-gap
PL is co-circular with respect to the excitation, having, under the
experimental conditions, maxima (white areas) at $\pi/4+n\pi$ and
minima (black areas) at $3/4\pi+n\pi$.\cite{note} Here $n=0,1$. The
opposite holds for \textit{i}-Ge and \textit{p}$^-$-Ge.

For $\mathrm{90 K\leq T\leq 170K}$, the direct-gap PL is still
circularly polarized. The helicity of photons emitted at the direct
gap transition in \textit{p}$^+$-Ge and \textit{n}-Ge, i.e. samples
with doping larger than $10^{16}$~$\mathrm{cm^{-3}}$, turns out to be
out of phase with respect to the one reported in the low temperature
regime for the same samples. On the other hand, the helicity of the
direct-gap emission in \textit{i}-Ge and \textit{p}$^-$-Ge is not
affected by the temperature variation, and it is co-circular with the
one of the doped samples. In this temperature regime, the angular
momentum of the direct-gap luminescence has the opposite direction as
the one of the absorbed photons at the excitation energy. In addition,
the amplitude of the peak modulation, and indeed the polarization
degree, is the largest for all the investigated samples.

Finally, in the high temperature range of Fig.~\ref{fig:3}(a),
i.e. above $\sim$170~K, the aforementioned sinusoidal pattern is
completely absent no matter the impurity content of the sample. This
result demonstrates that the emitted light is not circularly polarized
and that electron spins are no longer oriented prior to
recombination. We can therefore conclude that above 170~K there exists
a thermal activation of spin relaxation mechanisms. Such process
pertains to the material itself and not to the actual type and level
of doping.

The findings discussed above are further corroborated by the
temperature dependence of the polarization degree, $\rho$, of the
emitted light [see Fig.~\ref{fig:3}(b)]. For all the investigated
samples, $\rho$ has a bell-shaped structure and, in agreement with the
discussion above, $\rho$ approaches zero at high temperatures. It
should be noted that under the experimental conditions and according
to the definition given in Eq.~(\ref{eq:rho}), $\rho$ is negative for PL
being co-circular with the excitation.

The impact of doping on the polarization degree is elucidated in
Fig.~\ref{fig:3}(b) by the low temperature tail of $\rho$. Indeed at
$\mathrm{4}$~K, $\rho\sim -10\%$ for both \textit{n}-Ge and
\textit{p}$^+$-Ge, but it becomes positive when the impurity content
is decreased below $10^{15}$~$\mathrm{cm^{-3}}$, being few percent for
\textit{p}$^-$-Ge and reaching a maximum of $\sim +30\%$ in
\textit{i}-Ge. In \textit{p}$^+$-Ge and \textit{n}-Ge, $\rho$ changes
sign between 50~K and 90~K. Whereas in this temperature range $\rho$
is almost constant for \textit{i}-Ge and it slowly increases in
\textit{p}$^-$-Ge. Noteworthy, $\rho$ reaches a positive maximum
around 125~K for all the samples, being larger than 40\% in
\textit{i}-Ge. This is by far larger than the theoretical maximum of
25\% achievable in complete absence of spin relaxation mechanisms for
band-edge emission in direct-gap bulk materials.\cite{Zutic04}
Remarkably, such straightforward result can be obtained in Ge, as
opposed to III-V compounds, without applying external perturbations,
e.g. mechanical stress to remove VB degeneracy.

The puzzling dependence of $\rho$ upon temperature and doping,
reported in Fig.~\ref{fig:3}(b) for bulk Ge, does not have any
counterpart in the well-established literature dealing with direct-gap
semiconductors. Indeed, our experimental findings point out that
mechanisms related to subtleties of the Ge band structure might play
an important role in determining optical orientation as well as the
spin dynamics. To address this further, we introduce the following
theoretical analysis.

\section{Theory}
\label{sec:theo}

\begin{figure}
  \centering
  \includegraphics[width=8.6cm]{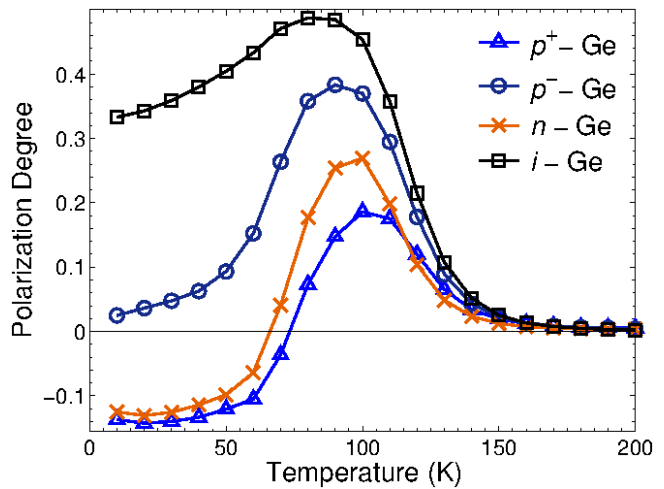}
  \caption{Calculated circular polarization degree of the
    recombination in the $\Gamma$ valley of bulk Ge excited with
    $\bm{\sigma}^-$.}
  \label{fig:theory}
\end{figure}

We use Monte Carlo simulations to interpret the direct-gap PL
in bulk Ge and to provide a solid, theoretical
framework for the experimental results in the previous
section. As shown in Fig.~\ref{fig:theory}, these simulations fully
recover the trends of the experiments for intrinsic, \textit{p}-type
and \textit{n}-type cases (without the use of fitting
parameters). Below we elaborate on implementation of the numerical
procedure and the description of the spin dynamics leading to the
polarization of the direct-gap PL.

To achieve accurate average of the direct-gap circular polarization
degree, we simulate $10^9$ photoexcited electrons. Each simulation
ends if the electron reaches the bottom of the $L$ valley or if it
experiences direct-gap radiative recombination while spending time in
the bottom of the $\Gamma$~valley.  Since the vast majority of
electrons relax to the bottom of the $L$ valley, the use of $10^9$
simulations guarantees accurate information on the small portion of
electrons that experience direct-gap radiative recombination. The spin
relaxation is not incorporated in these Monte Carlo simulations since
the simulated dynamics mimics the evolution of the electrons within
the first few ps. However, due to the cross-talks between the
populations of the different CB valleys, the high-temperature
luminescence is governed by thermally-activated electrons that reach
to the $\Gamma$ valley after spending a relatively long time in the
$L$ valleys (compared with the momentum, energy and spin relaxation
times).

To capture the radiative recombination from thermally-activated and
unpolarized electrons, the following considerations are taken. The
probability of photoexcited electrons to reside in the $L$ valley
rather than to undergo luminescence in the $\Gamma$ valley follows
$\eta_1\sim\tau_{\scriptscriptstyle{r,\Gamma}}/\tau_{\scriptscriptstyle{\Gamma\rightarrow
    L}}$, where $\tau_{\scriptscriptstyle{r,\Gamma}}\sim0.1$~ns and
$\tau_{\scriptscriptstyle{\Gamma\rightarrow L}}\sim1$~ps are
respectively the recombination time in the $\Gamma$ valley and the
scattering time from $\Gamma$ to the $L$ valley.\cite{Tanaka93} The
photoexcited electrons with density $N_e$ in steady state can also be
calculated from the laser intensity and the absorption coefficient,
which distribute mostly in the $L$ valley bottom where doped electrons
with density $N_d$ may also exist. We denote their ratio as
$\eta_2\equiv N_d/N_e$. Most of these electrons eventually recombine
with holes either in the $L$ valley or in the $\Gamma$ valley with
thermal activation. The rates of the former and latter are
respectively $\Gamma_a\sim\tau_{\scriptscriptstyle{r,L}}^{-1}$ and
$\Gamma_b\sim\tau_{\scriptscriptstyle{r,\Gamma}}^{-1}\exp(-\Delta\epsilon_{\scriptscriptstyle\Gamma,L}/k_BT)$,
where $\tau_{\scriptscriptstyle{r,L}}\sim1$~$\mu$s is the
recombination time in the $L$ valley. Hence with the circular
polarization degree of PL
$\rho_{\scriptscriptstyle{MC}}$ extracted from the Monte Carlo
simulations within the time scale of momentum relaxation, we can
derive the result with spin relaxation:
\begin{equation}
  \rho\approx\frac{\rho_{\scriptscriptstyle{MC}}}{1\!+\!\eta_1(1\!+\!\eta_2)\displaystyle\frac{\Gamma_b}{\Gamma_a\!+\!\Gamma_b}}\,.\label{eq:rho}
\end{equation}
The final results of Eq.~(\ref{eq:rho}) are depicted in
Fig.~\ref{fig:theory}. The dependence of the circular polarization
degree on temperature and doping reconciles with the empirical
results.  It is mentioned that the thermal-activation process of
unpolarized electrons form the $L$ valley rather some unrealistic
ultrafast spin relaxation in $\Gamma$ valley leads to the decay of the
circular polarization degree at temperatures above 100~K for all the
curves in Fig.~\ref{fig:theory}.  We now turn to the calculation
procedure of the ultrafast electron evolution following
photoexcitation from which we extract the value of
$\rho_{\scriptscriptstyle{MC}}$ in Eq.~(\ref{eq:rho}).

The photoexcitation is modeled by utilizing a pseudo-Voigt profile for
the CW Nd:YVO$_4$ laser.\cite{Wertheim74} In addition, momentum
alignment and spin-momentum correlation are found necessary to
generate the electron distributions immediately after the
photoexcitation.\cite{Qing11,Dymnikov76,Efanov83} The initial states
of electrons are extracted from an eight-band Kane model.\cite{Kane57}
For the momentum relaxation of photoexcited electrons, we incorporate
all sorts of intervalley electron-phonon scattering between different
$L$, $X$ or $\Gamma$ valleys, and also intravalley scattering within
each valley.\cite{Jacoboni81,Li12} In addition, we consider the
ionized impurity scattering, the carrier-carrier binary scattering,
and the collective electron-plasmon scattering.\cite{Lundstrom_MSSSD}
Below we elaborate on these details.

\subsection{Photoluminescence in Germanium}\label{MC_A}

(a) \textit{Band structure.} The CB of Ge includes four $L$
valleys, one $\Gamma$ valley and six $X$ valleys. In each valley, we
consider ellipsoidal constant energy surfaces as
$\gamma(\mathbf{k})=\epsilon(1+\alpha\epsilon)=\hbar^{2}(k_l^2/m_l+k_t^2/m_t)/2$,
where $\alpha$ is a nonparabolicity parameter,\cite{Conwell68} and
$\epsilon$ is the electron kinetic energy. $l$ ($t$) denotes the
longitudinal (transverse) components. We employ $m_l=1.588m_0$,
$m_t=0.0815m_0$, $\alpha=0.3$~eV$^{-1}$ in the $L$ valley,
$m_l=1.353m_0$, $m_t=0.288m_0$, $\alpha=0$ in the $X$ valley and
$m_l=m_t=0.037m_0$, $\alpha=0$ in the $\Gamma$
valley. \cite{Jacoboni81} We introduce the Herring-Vogt transformation
defined by $k_{i}^{*}=\mathcal{T}_{ij}^{n}k_{i}^{n}$.\cite{Herring56}
In the frame of reference of the $n^{th}$ valley, centered at the
bottom of the valley with the $z$ axis along its symmetry axis, we
take the transformation matrix
$\mathcal{T}^n=\mathrm{diag}\big(\sqrt{m_0/m_t},\sqrt{m_0/m_t},\sqrt{m_0/m_l}\,\big)$
so that the ellipsoidal constant energy surfaces become spheres. To
preserve vector equations, $\mathcal{T}^n$ is also applied to other
vector quantities such as phonon wavevectors.

The band parameters of Ge vary slightly with temperature and
doping. We adopt the dependence
$E_{g,\scriptscriptstyle\Gamma}(T)=E_{g,\scriptscriptstyle\Gamma}(0)-\alpha_{\scriptscriptstyle\Gamma}T^2/(T+\beta_{\scriptscriptstyle\Gamma})$
for the direct band gap,\cite{Varshni67} where
$E_{g,\scriptscriptstyle\Gamma}(0)=0.887$~eV,
$\alpha_{\scriptscriptstyle\Gamma}=5.82\times10^{-4}$~eV/K,
$\beta_{\scriptscriptstyle\Gamma}=296$~K, and also
$\Delta_{\scriptscriptstyle{SO}}=0.287$~eV for the split-off
energy. The band gap shrinkage takes 36~meV, 9~meV and 1~meV for
$p$-type $3.6\times10^{18}$~cm$^{-3}$ doping, $n$-type
$8.3\times10^{16}$~cm$^{-3}$ doping and $p$-type $10^{15}$~cm$^{-3}$
doping, respectively.\cite{Jain91}

\vspace{0.3cm} (b) \textit{Laser spectrum.} The center frequency
$\nu_c$ of the CW Nd:YVO$_4$ laser is located at
$h\nu_c=1.165$~eV. Considering the broadening effects and the
background noises, we assume the distribution of the photon energy
$E_l$ to have a pseudo-Voigt profile made of weighted contributions
from Gaussian and Lorentzian distributions,\cite{Thompson87}
\begin{equation}
  P(E_l)=\eta_{\scriptscriptstyle{G}}\,G(E_l;h\nu_c,\delta_{\scriptscriptstyle{G}})+(1\!-\!\eta_{\scriptscriptstyle{G}})L(E_l;h\nu_c,\delta_{\scriptscriptstyle{L}})\,,\nonumber
\end{equation}
where
\begin{eqnarray}
  G(E_l;h\nu_c,\delta_{\scriptscriptstyle{G}})&=&\frac{1}{\delta_{\scriptscriptstyle{G}}\sqrt{2\pi}}\exp\!\left[-\frac{(E_l\!-\!h\nu_c)^2}{2\delta_{\scriptscriptstyle{G}}^2}\right],\nonumber\\
  L(E_l;h\nu_c,\delta_{\scriptscriptstyle{L}})&=&\frac{1}{\pi}\!\left[\frac{\delta_{\scriptscriptstyle{L}}}{(E_l\!-\!h\nu_c)^2\!+\!\delta_{\scriptscriptstyle{L}}^2}\right].\nonumber
\end{eqnarray}
In the simulations we use $\eta_{\scriptscriptstyle{G}}=0.8$,
$\delta_{\scriptscriptstyle{G}}=1$~meV and
$\delta_{\scriptscriptstyle{L}}=6$~meV.

For a fixed $E_l$, electrons excited from different VB have different
initial energy. The corresponding density of states determines the
excited fraction. The total amount for each kind of excitation is
proportional to $(m_c/m_v+1)^{-3/2}\sqrt{E_l-E_{g,cv}}$ when
$E_l>E_{g,cv}$, or zero otherwise. Here $m_c$ and $m_v$ are the
effective masses of electrons and holes, respectively. The former
corresponds to $\Gamma$ valley, and the latter includes
$m_{hh}=0.28m_0$ for heavy holes, $m_{lh}=0.044m_0$ for light holes,
and $m_{so}=0.084m_0$ for split-off holes. $E_{g,cv}$ is the energy
gap between edges of the two bands in the photoexcitation process.

\begin{table}
  \caption{Momentum alignment and spin-momentum correlation for circular polarization at the $\Gamma$~point. $\hat{\mathbf{k}}$ is the unit vector in the direction of electron momentum and $\hat{\mathbf{p}}\equiv i\,\hat{\mathbf{e}}\times\hat{\mathbf{e}}^*$ is the photon angular momentum. For relatively small wavevector, these simple forms are good approximations. \cite{Qing11}}
  \vspace{0.1cm}
  \label{tab:SMC}
  \tabcolsep=0.2cm
  \renewcommand{\arraystretch}{0}
  \begin{tabular}{c|ccc}
    \hline \hline \\[-0.3cm]
    Band & Heavy-hole & Light-hole & Split-off \\[0.2cm] \hline \\[-0.2cm]
    $\!\!\!\begin{array}{c}\mbox{Momentum}\\\mbox{alignment}\end{array}\!\!\!$ & $\displaystyle\frac{3}{4}[(\hat{\mathbf{p}}\!\cdot\!\hat{\mathbf{k}})^2\!\!+\!1]$ & $\displaystyle\frac{5}{4}\!-\!\frac{3}{4}(\hat{\mathbf{p}}\!\cdot\!\hat{\mathbf{k}})^2$ & $\displaystyle1$ \\[0.5cm] $\!\!\!\begin{array}{c}\mbox{Spin-momentum}\\\mbox{correlation}\end{array}\!\!\!$ & $\displaystyle-
    \frac{(\hat{\mathbf{p}}\!\cdot\!\hat{\mathbf{k}})\hat{\mathbf{k}}}{1\!+\!(\hat{\mathbf{p}}\!\cdot\!\hat{\mathbf{k}})^2}$ & $\displaystyle\frac{3(\hat{\mathbf{p}}\!\cdot\!\hat{\mathbf{k}})\hat{\mathbf{k}}\!-\!2\hat{\mathbf{p}}}{5\!-\!3(\hat{\mathbf{p}}\!\cdot\!\hat{\mathbf{k}})^2}$ & $\displaystyle\frac{1}{2}\hat{\mathbf{p}}$ \\[0.5cm] \hline \hline
  \end{tabular}
\end{table}

\vspace{0.3cm} (c) \textit{Momentum alignment and spin-momentum
  correlation.} In momentum space at the instant of photoexcitation,
the alignment determines the number of electrons in each direction,
while the correlation assigns the corresponding average spin. Only at
the $\Gamma$~point, they possess simple analytical forms as shown in
Table~\ref{tab:SMC},\cite{Dyakonov_OO} where $\hat{\mathbf{e}}$ is the
unit polarization vector ($\hat{\mathbf{p}} =
i\hat{\mathbf{e}}\times\hat{\mathbf{e}}$). To generate realistic
values when electrons are far from the valley bottom, we utilize the
Kane Hamiltonian with spin-orbit coupling to calculate accurate
eigenvectors. They are formulated from the Luttinger parameters
$\gamma_1=13.35$, $\gamma_2=4.25$, $\gamma_3=5.69$ and the Kane matrix
element $E_P=26.3$~eV.\cite{Lew_kpM} To construct the density matrix
we first denote the eigenvectors of CB and VB respectively by
$|c_i\rangle$ and $|v_i\rangle$, where $i\in\lbrace1,2\rbrace$ in each
band, and $v\in\lbrace h,l,s\rbrace$ is comprised of heavy-hole,
light-hole and split-off bands. Then the coherent photoexcited states
are $\big|\alpha_i(v)\big\rangle \propto\sum_j\Big\lbrace\big\langle
c_j\big|\vec{\nabla}\big|v_i\big\rangle
\cdot\hat{\mathbf{e}}\Big\rbrace\big|c_j\big\rangle$, and the density
matrix is simply
$\mathcal{G}=\sum_i\big|\alpha_i(v)\big\rangle\big\langle\alpha_i(v)\big|$
for electrons excited from the corresponding band. We can transform it
to a $2\times2$ form with basis $\big|s\big\rangle$ and
$\big|s'\big\rangle$, namely $\mathcal{F}=\sum_{s,s'}\big|s\big
\rangle \big \langle s\big|\mathcal{G}\big|s'\big \rangle \big \langle
s'\big|$, where $s$ and $s'$ are spin indices.  Instead of what appear
in Table~\ref{tab:SMC}, the momentum alignment and spin-momentum
correlation respectively take the forms of $\mathrm{Tr}(\mathcal{F})$
and $\mathrm{Tr}(\hat{\bm{\sigma}}\mathcal{F})$, where
$\hat{\bm{\sigma}}$ denotes the Pauli matrix vector. Such numerical
results still bear notable similarities to the analytical forms at the
$\Gamma$~point, while they indeed capture some new features such as
warping effects (see supplemental material of
Ref.~[\onlinecite{Qing11}] for figures). We use the above analysis to
pre-generate tables for initial states in the simulation.

\begin{table*} \caption{ Coupling constants and phonon energies for
    intervalley scattering in Ge. These values are inherited from
    previous works,\cite{Jacoboni81,Jacoboni83} and can also be
    extracted from the empirical pseudopotential
    model.\cite{Chelikowsky76}} \vspace{0.1cm}
  \label{tab:D_Omg}
  \tabcolsep=0.39cm
  \renewcommand{\arraystretch}{1.5}
  \begin{tabular}{c|ccccccc}
    \hline \hline
    Scattering & $XL$ (LA) & $\Gamma X$ (LA) & $\Gamma L$ (LA) & $XX_g$ (LA) & $XX_g$ (LO) & $LL$ (LA, LO) & $LL$ (TA) \\ \hline \hline
    $D_{iv}$ [eV/\AA] & 10.0 & 4.06 & 2.0 & 0.789 & 9.46 & 3.0 & 0.20 \\
    $\Omega_{iv}$ [meV] & 27.6 & 27.6 & 27.6 & 8.62 & 37.1 & 27.6 & 10.3 \\ \hline \hline
  \end{tabular}
\end{table*}

\vspace{0.3cm} (d) \textit{Luminescence.} It is considered when the
relaxed electrons approach the valley bottom. The proper range can be
estimated from the width of the direct-gap radiation peak in the
PL spectra. We choose 12~meV, close to the full width
at half maximum. As the electron kinetic energy drops below, radiation
times are randomized according to a homogeneous Poisson process,
$\propto\exp(-t/\tau_{\scriptscriptstyle{r,\Gamma}})$. At the time of
recombination, the density matrix of an electron with spin
$\mathbf{S}$ can be reconstructed as
$\mathcal{F}=[\mathcal{I}+\hat{\bm{\sigma}}\cdot\mathbf{S}]/2$.

With the aforementioned coherent state
$\big|\alpha_i(v)\big\rangle\propto\sum_j\Big\lbrace \big\langle
c_j\big|\vec{\nabla}\big|v_i\big\rangle\cdot\hat{\mathbf{e}}'\Big\rbrace\big|c_j\big\rangle$,
we can calculate the total intensity with polarization
$\hat{\mathbf{e}}'$ as follows,\cite{Dyakonov_OO}
\begin{equation}
  I_{\hat{\mathbf{e}}'}=\!\int\!d^3\mathbf{k}\sum_i\big\langle\alpha_i(v)\big|\mathcal{F}\big|\alpha_i(v)\big\rangle\,.\nonumber
\end{equation}
Here $\big|c_i\big\rangle$ and $\big|v_i\big\rangle$ approximate to
the simple wave functions at $\Gamma$~point. The circular polarization
degree is defined as $\rho=(I_+-I_-)/(I_++I_-)$, where $I_+$ and $I_-$
are the intensities of right and left circularly polarized
radiation. Especially, for thermalized electrons with no spin-momentum
correlation, we find that $\rho=-\mathbf{S}\cdot\hat{\mathbf{n}}'$
holds for the recombination with both heavy and light holes. Here
$\mathbf{S}$ is the average spin and $\hat{\mathbf{n}}'$ is the unit
vector in the observation direction of the luminescence.

\subsection{Intrinsic Scattering Mechanisms}\label{MC_B}

(a) \textit{Intervalley phonon scattering.} After integrating over all
possible states in the final valley, the total scattering rate reads
\begin{equation}
  \renewcommand{\arraystretch}{0.5}
  \Gamma_{iv} =\frac{D_{iv}^2m_{d(f)}^{3/2}Z_f}{\sqrt{2}\pi\hbar^2\varrho\,\Omega_{iv}}\,\Phi_{iv}\!\left(\!\!\begin{array}{c}n_{\mathbf{q}}\sqrt{\gamma\!\left(\epsilon\!+\!\Omega_{iv}\!-\!\Delta\epsilon_{fi}\right)}\\\\(n_{\mathbf{q}}\!+\!1)\sqrt{\gamma\!\left(\epsilon\!-\!\Omega_{iv}\!-\!\Delta\epsilon_{fi}\right)}\end{array}\!\!\right)\!,\label{eq:Gm_iv}
\end{equation}
Here and in Eq.~(\ref{eq:Gm_ac}) the top (bottom) line refers to phonon
absorption (emission). $m_{d}=\sqrt[3]{m_lm_t^2}$ is the
density-of-states effective mass, and `$(f)$' denotes the final
valley. $\varrho=5.32$~g/cm$^3$ and $n_{\mathbf{q}}$ are the crystal
density and Bose-Einstein distribution,
respectively. $\Phi_{iv}=1+2\alpha(\epsilon\pm\Omega_{iv}-\Delta\epsilon_{fi})$
is the extra factor due to the nonparabolicity, where $\pm$ holds for
top or bottom line and the same in $\Phi_{ac}$ for
Eq.~(\ref{eq:Gm_ac}). $\Delta\epsilon_{fi}$ is the energy difference
of the final and initial valley bottoms. We characterize Ge with
$\Delta\epsilon_{\scriptscriptstyle{X,L}}=0.18$~eV,
$\Delta\epsilon_{\scriptscriptstyle{X,\Gamma}}=0.04$~eV, and
$\Delta\epsilon_{\scriptscriptstyle{\Gamma,L}}=0.14$~eV. The coupling
constant $D_{iv}$ and the phonon energy $\Omega_{iv}$ of the
corresponding mechanism are listed in Table~\ref{tab:D_Omg}. $Z_f$ is
the number of possible final valleys. The equivalent final valley is
chosen randomly. The states after scattering are equally probable on
the constant energy surface with energy conservation.

\begin{table}[b]\caption{
Integration limits in Eq.~(\ref{eq:Gm_ac}). They guarantee the energy and momentum conservation. (See supplemental material of Ref.~[\onlinecite{Li12}])}
\vspace{0.1cm}
\label{tab:x1_x2}
\tabcolsep=0.25cm
\renewcommand{\arraystretch}{1.5}
\begin{tabular}{ccc}
  \hline \hline
  Condition & Absorption & Emission \\ \hline
  $\epsilon\le\epsilon_s$ ($x_\epsilon\le x_0$) & $x_1=x_0-\sqrt{x_0x_\epsilon}$ & absent \\
  & $x_2=x_0+\sqrt{x_0x_\epsilon}$ & \\ \hline
  $\epsilon>\epsilon_s$ ($x_\epsilon>x_0$) & $x_1=0$ & $x_1=0$ \\
  & $x_2=\sqrt{x_0x_\epsilon}+x_0$ & $x_2=\sqrt{x_0x_\epsilon}-x_0$ \\ \hline \hline
\end{tabular}
\end{table}

\vspace{0.3cm} (b) \textit{Intravalley optical phonon scattering.}
This effect is relatively weak in crystals with inversion
symmetry. Yet we formalize it in the same way we did for the
intervalley phonon scattering [when $D_{op}$ and $\Omega_{op}$
replace, respectively, $D_{iv}$ and $\Omega_{iv}$ in
Eq.~(\ref{eq:Gm_iv})]. Meanwhile, $Z_f=1$ and
$\Delta\epsilon_{fi}=0$. For electrons in $L$ valley, we consider
$D_{op}=5.5$~eV/\AA, $\Omega_{op}=37.1$~meV.

\vspace{0.3cm} (c) \textit{Intravalley acoustic phonon scattering.} We
take the long wavelength approximation without distinguishing between
longitudinal and transverse phonons. With the dimensionless variable
\begin{equation}
  x=\frac{\hbar qv_s}{k_BT}\approx\frac{\hbar q^*v_s }{k_BT}\sqrt{\frac{m_d}{m_0}}\,,\label{eq:x_q}
\end{equation}
where $\mathbf{q}=\mathbf{k}-\mathbf{k}'$, the total scattering rate
follows
\begin{equation}
  \renewcommand{\arraystretch}{0.7}
  \Gamma_{ac}=\frac{\Xi^2m_d^{1/2}(k_BT)^3}{2\sqrt{2}\pi\hbar^4\varrho\,v_s^4}\frac{1}{\sqrt{\gamma(\epsilon)}}\!\int^{x_2}_{x_1}\!\!\Phi_{ac}\!\left(\!\begin{array}{c}n_\mathbf{q}\\\\n_\mathbf{q}\!+\!1\end{array}\!\right)\!x^2dx\,,\label{eq:Gm_ac}
\end{equation}
where $v_s=5.4\times10^5$~cm/s denotes the sound velocity. The
deformation potential $\Xi$ in $L$, $X$ and $\Gamma$ valley has values
of 11~eV, 9~eV and 5~eV, respectively. The integration limits $x_1$
and $x_2$ in Eq.~(\ref{eq:Gm_ac}) are given in Table~\ref{tab:x1_x2}
where $\epsilon_s=m_d\,v_s^2/2$, $x_\epsilon=4\epsilon/k_BT$,
$x_0=4\epsilon_s/k_BT$.\cite{Li12}
$\Phi_{ac}=1+2\alpha\epsilon\pm2\alpha k_BTx$ is the nonparabolic
factor. With the truncated Laurent expansion of phonon distribution
\begin{equation}
  \renewcommand{\arraystretch}{1.3}
  n_\mathbf{q}(x)=\left\{\begin{array}{ll}
      1\big/x\!-\!\frac{1}{2}\!+\!\frac{1}{12}x\!-\!\frac{1}{720}x^{3}\!+\!\frac{1}{30240}x^{5}, & \ \text{if \ensuremath{x<4}\,,}\\
      0, & \ \text{if \ensuremath{x\geq4}}\,,
    \end{array}\right.\nonumber
\end{equation}
the integration in Eq.~(\ref{eq:Gm_ac}) can then be readily performed.
For the state after scattering, we choose $x$ according to the
expression inside the integration in Eq.~(\ref{eq:Gm_ac}) with the
rejection technique.\cite{Jacoboni83} Then $q^*$ is calculated from
Eq.~(\ref{eq:x_q}), and the angle between $\mathbf{k}'^{*}$ and
$\mathbf{k}^*$ is obtained by energy and momentum conservation. The
angle of rotation around $\mathbf{k}^*$ is completely random.

\vspace{0.3cm} (d) \textit{Rees self-scattering.} To determine the
occurrence of scattering events without the difficulty of solving
integral equations for each event, we use the imaginary
self-scattering technique.\cite{Rees68, Rees69} A homogeneous Poisson
process of scattering is simulated with the rate parameter $\Gamma_0$,
in which a fictitious self-scattering is included such that the total
scattering rate together with the self-scattering is $\Gamma_0$. If an
electron undergoes a `self-scattering' event, its wavevector
immediately before and after the scattering is unchanged. This
technique is valid if $\Gamma_0$ exceeds the total scattering rate of
an electron in state $\mathbf{k}$. In the simulation, we use
$\Gamma_0=10^{15}$~s$^{-1}$.

\subsection{Scattering Mechanisms in Doped Samples}\label{MC_C}

(a) \textit{Ionized impurity scattering.} Using the Brooks-Herring
approach,\cite{Jacoboni83} we get a total scattering rate for ionized
impurity with density $N_I$:
\begin{equation}
  \Gamma_{I}=\frac{\sqrt{2}e^4N_Im_d^{3/2}}{\pi\hbar^4\varepsilon_0^2\varepsilon_r^2\beta^4}\sqrt{\gamma(\epsilon)}\,\Phi_{I}\!\left(\!1\!+\!\frac{8m_d\gamma(\epsilon)}{\hbar^2\beta^2}\!\right)^{\!\!-1}\!\!\!\!,\label{eq:Gm_I}
\end{equation}
where $\varepsilon_r=16$ is the relative dielectric constant, and
$\Phi_I=1+2\alpha\epsilon$ is the nonparabolic factor. The screening
$\beta^{-1}$ is taken as the Debye length
\begin{equation}
  \beta^{-1}=L_D=\sqrt{\frac{\varepsilon_0\varepsilon_r k_BT}{N_de^2}}\,.\label{eq:bt}
\end{equation}
By randomizing a number $r\in[0, 1]$, the scattering angle $\theta$
follows
\begin{equation}
  \cos{\theta}=1\!-\!2(1\!-\!r)\!\left(\!1\!+\!\frac{8m_d\gamma(\epsilon)r}{\hbar^2\beta^2}\!\right)^{\!\!-1}\!\!\!\!.\nonumber
\end{equation}

\vspace{0.3cm} (b) \textit{Carrier-carrier scattering.} The
interaction is between photoexcited electrons and a Fermi-Dirac
distribution of thermal carriers, $f(\epsilon)$, due to the background
doping. We consider a screened Coulomb potential where the total
scattering rate is derived from,\cite{Dewey93,Dery03}
\begin{equation}
  \Gamma_{cc}=\frac{e^4N_dm_0^{3/2}}{2\pi\hbar^3\varepsilon_0^2\varepsilon_r^2\mu_d^{1/2}\beta^{*2}}\!\int\!d^3\mathbf{k}_s^*\frac{\big|\mathbf{g}^*\big|}{\beta^{*2}\!+\!\mathbf{g}^{*2}}f[\epsilon(\mathbf{k}_s^*)]\,.\label{eq:Gm_cc}
\end{equation}
Here $\beta^*=\beta\sqrt{\mu_d/m_0}$ and
$\mu_d=(\mu_x\mu_y\mu_z)^{1/3}$ where
$2\mu_i^{-1}=m_i^{-1}+m_{i(s)}^{-1}$ and $i\in\{x,y,z\}$. Parameters
of thermal carriers are denoted by $(s)$, which in the case of
$p$-type and $n$-type Ge correspond to heavy holes and
$L$$\,$-$\,$valley electrons,
respectively. $g_i^*=\sqrt{m_0\mu_i}[k_i/m_i-k_{s,i}/m_{i(s)}]$ is the
transformed relative wavevector, and $\mathbf{k}_s$ is the wavevector
of a thermal carrier. The integration can be evaluated numerically,
and we pre-generate a table for the simulation. After a specific
scattering event, $\mathbf{k}^{\prime*}$ is determined by
\begin{equation}
  k_i^{\prime*}=\frac{1}{2}m_i\!\left(\!\frac{k_i^*}{m_i}\!+\!\frac{k_{s,i}^{*}}{m_{i(s)}}\!-\!\frac{g_i^{\prime*}}{\mu_i}\!\right),\nonumber
\end{equation}
where $\mathbf{k}_s$ is generated randomly from the Fermi-Dirac
distribution. We notice
$\big|\mathbf{g}^{\prime*}\big|=\big|\mathbf{g}^*\big|$, and the angle
between $\mathbf{g}^{\prime*}$ and $\mathbf{g}^*$ is given by
\begin{equation}
  \cos{\vartheta}=1\!-\!2(1\!-\!r)\!\left(\!1\!+r\frac{\mathbf{g}^{*2}}{\beta^{*2}}\!\right)^{\!\!-1}\!\!\!\!,\nonumber
\end{equation}
with a random number $r\in[0, 1]$. The azimuthal angle has no
preferences from 0 to $2\pi$.

The screening takes the form of Eq.~(\ref{eq:bt}) only when the
energies of involved carriers are close to each other. Thus we only
use Eq.~(\ref{eq:Gm_cc}) when the energy of the photoexcited electron
is slightly (6~meV) above the defined valley bottom [within $12$~meV,
see Sec.~\ref{MC_A}(d)] plus the mean energy of the doped
carriers. This limitation is removed if the plasmon scattering is not
effective (see below). In addition, from the analysis of
Eq.~(\ref{eq:Gm_cc}) we note that when the photoexcited electron is in
the $\Gamma$ valley, its much smaller effective mass compared with
that of the background carrier renders the binary collision process
ineffective in relaxing the photoexcited electron to the bottom of the
valley (intervalley phonon-induced relaxation mechanisms become
faster). The binary process is more effective in the $L$ and $X$
valleys where the effective masses are comparable with those of the
background carriers.

\vspace{0.3cm} (c) \textit{Plasmon scattering.} With the plasmon-pole
approximation,\cite{Haug_TOEPS,Mahan_MPP} we can calculate the total
scattering rate
\begin{equation}
  \Gamma_{pl}=\frac{e^3N_d^{1/2}m_d^{1/2}}{4\sqrt{2}\pi\hbar(\varepsilon_0\varepsilon_r)^{3/2}m_{d(s)}^{1/2}}\frac{1}{\sqrt{\gamma(\epsilon)}}\,\Phi_{pl}\!\int_{q_1^*}^{q_2^*}\!\frac{1}{q^*}dq^*\,,\label{eq:Gm_pl}
\end{equation}
where the antiderivative of the integrand is simply a logarithmic
function. The nonparabolic factor is
$\Phi_{pl}=1+2\alpha\epsilon$. The integral interval $[q_1^*,q_2^*]$
satisfies
\begin{equation}
  \left|\frac{1}{2q^*\sqrt{2m_d\gamma(\epsilon)}}\bigg(\!2m_d\omega_{pl}\sqrt{1\!+\!\frac{q^{*2}}{\beta^2}}\!-\!\hbar q^{*2}\!\bigg)\right|\leq1\,,\label{eq:pl_judge}
\end{equation}
where $\omega_{pl}
=(N_de^2\big/m_{d(s)}\varepsilon_0\varepsilon_r)^{1/2}$ is the plasma
frequency. For the state after scattering, we choose $q$ according to
the distribution inside the integration in Eq.~(\ref{eq:Gm_pl}). The
angle between $\mathbf{q}^*$ and $\mathbf{k}^*$ is determined by
energy conservation, in which the electron loses the energy of
$\hbar\omega_{pl}\sqrt{1+(q^*/\beta)^2}$, and the azimuthal angle is
completely random. Eventually, we obtain
$\mathbf{k}^{\prime*}=\mathbf{k}^{*}+\mathbf{q}^{*}$.

We notice that Eq.~(\ref{eq:pl_judge}) is not always possible,
especially when $\epsilon\lesssim\hbar\omega_{pl}$. In this case, we
broaden the suitable range of the binary carrier-carrier scattering to
avoid discontinuity. We note that the restriction
$\epsilon\lesssim\hbar\omega_{pl}$ can be readily achieved in the
$\Gamma$~valley due to its small effective mass (compared with the $L$
and $X$ valleys).

\section{Physical Picture}
\label{sec:physpic}

Here we highlight the underlying physics leading to the results
outlined in the previous experimental and theoretical sections.
Following the photoexcitation of spin-polarized electrons in the
$\Gamma$~valley, their vast majority relax to the bottom of the $L$
valleys, located $\Delta\epsilon_{\scriptscriptstyle\Gamma,L} \simeq
0.14$~eV below the edge of the $\Gamma$ valley. The lifetime of
electrons in the $L$~valley is relatively long ($>$$\,$1~$\mu$s) and
governed by non-radiative recombination channels across the indirect
band gap. Our work, however, focus on the minute fraction of electrons
that experience direct-gap radiative recombination from the bottom of
the $\Gamma$~valley. The circular polarization degree of this
luminescence provides useful information on the dominant relaxation
mechanisms of photoexcited electrons.


The results from experiments and theory summarized in
Fig.~\ref{fig:3}~(b) and Fig.~\ref{fig:theory} show a strong decay of
the circular polarization degree at high temperatures. We can infer
that radiative recombination of unpolarized electrons already plays a
role at 120~K and it dominates the luminescence above 170~K. The
source of these unpolarized electrons is from thermal-activation of
$L$ valley electrons. Specifically, electrons from the bottom of the
$L$ valleys can visit the $\Gamma$~valley with a probability of $\sim
\exp(-\Delta\epsilon_{\scriptscriptstyle\Gamma,L}/k_BT)$. During these
sporadic visits there is a minute probability for the electrons to
experience radiative recombination rather than phonon-induced
ultrafast scattering back to the $L$~valley. Given that the
recombination lifetime of electrons in the $L$ valley is longer than
1~$\mu$s and that their spin relaxation time is of the order of 1~ns
at high temperatures,\cite{Li12b} the majority of the ultrashort
visits in the $\Gamma$~valley take place when the electrons are no
longer spin-polarized.

Below $\sim$120$\,$K, the aforementioned thermal-activation process is
practically quenched, and the luminescence has contributions from two
types of spin-polarized electron populations. The first is of
electrons excited from the SO, which results from the finite energy
width of the laser excitation (see Sec. \ref{MC_A}). These electrons
are excited at the bottom of the $\Gamma$ valley and in what follows
we term them `low-energy' electrons. As expected from the selection
rules, \cite{Zutic04} upon radiative recombination they provide a
circular polarization degree of +50\%. The second population is of
electrons excited from heavy and light hole VB. These electrons need
to relax more than 100~meV before reaching the bottom of the $\Gamma$
valley and in what follows we term them `high-energy' electrons. Only
after this energy relaxation, their average contribution of $-$25\% to
the circular polarization degree \cite{Zutic04} can be noticed in the
peak of the direct band gap PL.

The relative contributions of `low-energy' and `high-energy' electron
populations to the direct-gap luminescence strongly depend on
excitation, temperature and doping conditions. Figure~\ref{fig:3}~(b)
and Fig.~\ref{fig:theory} show that below 50$\,$K the circular
polarization degrees are relatively constant and their
doping-dependent values range from $\sim -$10\% in $p^{+}$-Ge to
nearly $\sim +$30\% in \textit{i}-Ge. In between 50$\,$K and 120$\,$K,
the circular polarization degree of all samples increases reaching
nearly +50\% in the intrinsic sample. The change in behavior around
$\approx 50$~K is attributed to the excitation conditions, as discussed in
Sec. \ref{subsec:PL}.  The density of `low-energy' electrons can
significantly increase at higher temperatures because of the band gap
shrinkage and because carrier-carrier interactions become
relevant. This increase explains the behavior of the circular
polarization degree above 50~K in all samples, where a value of +50\%
indicates that only these electrons contribute to the direct band gap
luminescence. At temperatures below 50$\,$K, the density of
`low-energy' electrons is low and relatively constant given the fact
that the energy band gap shrinks only by $\sim$3~meV from zero to
50$\,$K. See Sec.~\ref{MC_A} for the analysis of the laser line and
excitation conditions.

The density of `high-energy' electrons reaching the bottom of the
$\Gamma$ valley after relaxation strongly depends on doping
conditions. In the intrinsic sample, the energy relaxation of
`high-energy' electrons is governed by phonon-induced intervalley
scattering between $\Gamma$, $X$ and $L$ valleys where after each
phonon emission the electron lose a few tens of meV. Since it is
highly likely for `high-energy' electrons to be transferred out of the
$\Gamma$ valley during such energy relaxation, the direct-gap
luminescence is governed by `low-energy' electrons. This physics
explains the positive and high circular polarization degree from the
intrinsic sample at low temperatures (in spite of the relatively small
population of `low-energy' electrons at these temperatures). In the
doped samples, on the other hand, the energy relaxation of
`high-energy' electrons is governed by collisions with the background
carriers. These Coulomb collisions come from a binary process in which
a photoexcited electron collides with a background
carrier,\cite{Dewey93,Dery03} and also a collective process in which
the photoexcited electron interacts with the thermal plasma of
background carriers.\cite{Haug_TOEPS,Mahan_MPP}

The energy relaxation of photoexcited electrons due to Coulomb
collisions is effective in the $X$ and $L$ valleys but not in the
$\Gamma$ valley. This difference stems from the effective mass as
explained in Sec.~\ref{MC_C}. A typical scenario at low temperatures
is that photoexcited electrons in the $\Gamma$~valley experience
phonon-induced intervalley scattering to one of the $X$ valleys since
it has higher rate compared with scattering to the $L$
valleys.\cite{Zhou94} The hot electrons then thermalize to the bottom
of the $X$~valley via the plasmon and carrier-carrier scattering
mechanisms.  The phonon-induced intervalley scattering typically takes
place after electrons reach the bottom of the $X$ valley transferring
some of the electrons back to the $\Gamma$ valley whereas most scatter
to the $L$ valley. Since $\Delta\epsilon_{X,\scriptscriptstyle\Gamma}
\simeq 40$~meV, electrons that scatter back to the $\Gamma$~valley
reach the valley bottom where they can effectively contribute to the
direct band gap luminescence. These characteristics lead to the
similarity between the curves of the fairly doped samples, despite of
the big gap in the doping level of $n$-Ge and $p^{+}$-Ge, namely
$8.3\times10^{16}$~cm$^{-3}$ and $3.6\times10^{18}$~cm$^{-3}$. For
$p^-$-Ge ($10^{15}$~cm$^{-3}$), the plasmon scattering is negligible,
and the binary carrier-carrier scattering is not as efficient at low
temperatures. Its circular polarization degree is therefore located in
between the heavily-doped and intrinsic samples, as expected.

Finally, we note that the cooling process in the $X$ valley mediated
by background impurities is expected to lead to an increase of the
direct-gap emission with the doping level. Such a general result has
to take place in both $n$- and $p$-doped Ge samples. In the former,
this contribution sums up to the higher electron population due to the
increased Fermi level with $n$-type doping.\cite{Sun09}

\section{Conclusions}
\label{sec:con}

In this work we have reported a joint experimental and theoretical
study of the polarization of light emitted in transitions across the
direct band gap of bulk Ge. The optical investigation, based upon
polarimetric analysis, provides compelling confirmation that optical
orientation of carrier spin can be achieved in bulk Ge.  The detailed
analysis of the state of light polarization further demonstrates that
the degree and helicity of direct-gap emission remarkably depend upon
parameters such as doping and temperature.

We emphasize that by combining continuous-wave polarization-resolved
PL experiments and Monte Carlo calculations, we gather simultaneous
information about spin and carrier relaxation mechanisms, without
relying on time resolved techniques. Indeed, in the optical
orientation process, electrons are photo-generated in the
$\Gamma$ valley from optically coupled heavy hole or split-off valence
band states at different energies and with opposite spin
orientation. Such an information does not get wiped out during the
subsequent radiative recombination, because, owing to the ultrafast
lifetime of $\Gamma$ electrons,\cite{Zhou94} it remains encoded in the
angular momentum of the emitted light. As a result, the measurement of
the helicity and polarization degree of the band-edge luminescence
allows us to keep track of the dominant non-equilibrium spin-polarized
population of electrons experiencing radiative recombination, and to
infer important information about energy and spin relaxation
channels. Thermal activation of unpolarized $L$ valley electrons is
shown to explain the luminescence depolarization at high temperatures,
whereas the doping level accounts for the different state of light
polarization in the low temperature regime. Finally, our findings
point out the pivotal impact in the cooling process of hot electrons
played by carrier-plasmon scattering within the $X$ valleys, whose
role in defining carrier dynamics has been largely overlooked in many
of the previous literature works dealing with direct-gap luminescence
in Ge.

The study of the injection of spin-polarized carriers and of
circularly polarized emission at the direct-gap of Ge is important
also application wise. We anticipate that a systematic calibration in
the low temperature regime of the dependence of the polarization
degree $\rho$ upon doping can possibly lead to the development of a
diagnostic tool for the determination of the impurity content in
Ge. Such spectroscopic method is non-destructive and has a high
spatial resolution ($\approx \mu m$). In addition, it does not require
accurate sample preparation and can be applied to address a doping
range not easily accessible by conventional techniques, e.g. Energy
Dispersive X-ray spectroscopy.

Finally, the recently discovered lasing action in Ge on Si
heterostructures,\cite{Liu10, Camacho12} albeit
debated,\cite{Carroll12, Boztug12, Dutt12, Wang13} holds the promise
of laser sources monolithically integrated onto the mainstream CMOS
platform,\cite{Liang10} thus filling the gap for the development of
the active devices needed to ground Si-photonics. At present, however,
direct-gap electroluminescence\cite{Hu09, Chaisakul11,Gallacher12,
  Nam12} and lasing\cite{Camacho12} in Ge-based heterostructures have
been achieved only under high current densities and shown to be not
efficient yet.\cite{Dutt12, Nam12} Such drawbacks strongly hamper the
widespread application of Ge-based light sources.

We suggest, however, that pursuing the concept of spin-based Ge
emitters can provide a sizable improvement in this
field.\cite{Rudolph03, Holub07, Holub11, Lee12} Pumping spin-polarized
carriers in the optical gain medium by using spin selective contacts
or circularly polarized light can reduce the injection threshold
required for electroluminescence or lasing action, finally boosting
the performances of the Ge emitters and their implementation into
Si-photonics circuits.

\begin{acknowledgments}
  The authors are grateful to F. Bottegoni and F. Ciccacci for
  fruitful discussions. We thank S. Bietti and S. Sanguinetti for the
  loan of one of the studied sample and R. Sorrentino and A. Colombo
  for technical assistance with the measurements. The theory work at
  UR is supported by NSF Contract No. ECCS-1231570 and by DTRA
  Contract No. HDTRA1-13-1-0013. The experimental work is supported by
  the Italian ministry MIUR through the PRIN Project
  No. 20085JEW12. F.P. acknowledges the support from Regione Lombardia
  via the Grant: Dote ricercatori.
\end{acknowledgments}



\end{document}